\documentclass[12pt]{article}
\pdfoutput=1
\usepackage{putex}

\usepackage{hyperref}

\usepackage{graphicx}
\usepackage{epstopdf}
\usepackage{enumerate}
\usepackage{cite}
\usepackage{tensor}
\usepackage{slashed}

\numberwithin{equation}{section}

\newcommand {\be} {\begin {equation}}
\newcommand {\ee} {\end {equation}}

\newcommand {\bes} {\begin {equation*}}
\newcommand {\ees} {\end {equation*}}

\newcommand{\es}[2] {\begin{equation} \label{#1} \begin{split} #2 \end{split} \end{equation}}

\newcommand{\beq}{\begin{equation}}
\newcommand{\eeq}{\end{equation}}

\begin{document}

\preprint{PUPT-2420}

\institution{PU}{Department of Physics, Princeton University, Princeton, NJ 08544}

\authors{Benjamin R.~Safdi}

\title{Exact and Numerical Results on Entanglement Entropy in $(5+1)$-Dimensional $\text{CFT}$ }

\abstract{

We calculate the shape dependence of entanglement entropy in $(5+1)$-dimensional conformal field theory in terms of the extrinsic curvature of the entangling surface, the opening angles of possible conical singularities, and the conformal anomaly coefficients, which are required to obey a single constraint.  An important special case of this result is given by the interacting $(2,0)$ theory describing a large number of coincident M5-branes.  To derive the more general result we rely crucially on the holographic prescription for calculating entanglement entropy using Lovelock gravity.  
We test the conjecture by relating the entanglement entropy of the free massless $(1,0)$ hypermultiplet in $(5+1)$-dimensions to the entanglement entropy of the free massive chiral multiplet in $(2+1)$-dimensions, which we calculate numerically using lattice techniques.  We also present a numerical calculation of the $(2+1)$-dimensional renormalized entanglement entropy for the free massive Dirac fermion, which is shown to be consistent with the $F$-theorem.               

}

\maketitle

\tableofcontents

\section{Introduction} 

The ground state entanglement entropy (EE) in quantum field theory measures the quantum correlations between two subspaces separated by a surface $\Sigma$, called the entangling surface (see, for example,~\cite{Eisert:2008ur,Nishioka:2009un,Calabrese:2009qy,Casini:2009sr,Solodukhin:2011gn,tadashi}).  More precisely, if $\rho = | 0 \left> \right< 0 | $ is the density matrix of the ground state, then we are to construct the reduced density matrix $\rho_R$ obtained by tracing over the degrees of freedom within one of the subspaces.  The entanglement entropy is defined as the Von Neumann entropy of the reduced density matrix: $S \equiv - \tr( \rho_R \log \rho_R)$.  The entanglement entropy is dominated by short-range correlations across the entangling surface.  If $\epsilon$ is the short-distance cut-off of the $(d+1)$-dimensional field theory, these correlations give a contribution to the entanglement entropy proportional to the area of $\Sigma$ divided by $\epsilon^{d-1}$.         

The subleading terms in the entanglement entropy contain useful, cut-off independent information about the field theory.  One application of these terms has been to characterize the topological order of the gapped ground state of many-body systems~\cite{Kitaev:2005dm,Levin:2006zz}.  Another application was to propose the $F$-theorem in $(2+1)$-dimensional QFT~\cite{Myers:2010tj,Jafferis:2011zi,Casini:2011kv,Klebanov:2011gs}, which states that the negative of the finite part of the entanglement entropy decreases under RG flow from a UV fixed point to an IR fixed point.  In even dimensional CFT it is expected that the entanglement entropy across a smooth entangling surface contains a $\log \epsilon $ divergence, with the coefficient of the term related to the conformal anomaly.  In odd dimensions there is no conformal anomaly, and as a result when the entangling surface $\Sigma$ is a smooth submanifold the entanglement entropy is not expected to have a $\log \epsilon$ divergence.  One way to see how the $\log \epsilon$ divergence arises is to consider the replica trick for calculating entanglement entropy.  In this approach, the entanglement entropy across a surface $\Sigma$ is related to the partition function of the theory on a copy of the background spacetime which is conically  singular along the entangling surface.

The form of the $\log \epsilon$ divergence in the entanglement entropy is known in $2$ and $4$-spacetime dimensions.  In $(1+1)$-dimensional CFT  
the entanglement entropy across an interval of length $L$ takes the simple form $S_L = - (c/3) \log \epsilon + O(\epsilon^0)$, where $c$ is the central charge.  In $(3+1)$-dimensional CFT there are two Weyl anomaly coefficients\footnote{We normalize $a$ and $c$ so that they both equal one for the real scalar.} $a$ and $c$.  With the entangling surface $\Sigma$ embedded in $\mathbb{R}^{1,3}$, the CFT entanglement entropy is given by~\cite{Solodukhin:2008dh}:
\es{threegen}{S_\Sigma= \alpha {A_\Sigma \over \epsilon^2}+
 \left ( \frac{a}{180} \int_\Sigma  \, E_2 + \frac{c}{240 \pi} \int_{\Sigma} \,( \tr k^2 - \frac{1}{2} k^2 )\right ) \log \epsilon + O(\epsilon^0) \,,
 }
 where $E_2$ is the Euler density on $\Sigma$ normalized such that $\int_{S^2} \, E_2 = 2$, and $k^i_{ab}$ is the extrinsic curvature, with ${a,b, ...}$ local indices on $\Sigma$ and $i=0,1$ labeling the unit normal vectors $n^i_{\mu}$, one of which is timelike and the other spacelike.  We use the standard notation $k^2 \equiv \gamma^{ab}  \gamma^{cd} \eta_{ij} k^i_{ab} k^j_{ cd}$ and $\tr k^2 \equiv  \gamma^{ad} \gamma^{bc} \eta_{ij} k^i_{ab} k^{j}_{cd}$, with $\gamma_{ab}$ the induced metric on $\Sigma$ and $\eta_{ij}$ the 2-dimensional Minkowski metric.  The functionals of extrinsic curvature which multiply $a$ and $c$ are invariant under diffeomorphisms on $\Sigma$ and under conformal transformations.    
 
 One of our main results is to find the form of the $\log \epsilon$ divergent term in the entanglement entropy in $(5+1)$-dimensional CFT in flat Minkowski space when the smooth entangling surface $\Sigma$ is embedded in a constant-time hyperplane ($k^0_{ab} = 0$).  There are 4 Weyl anomaly coefficients $(A,B_1,B_2,B_3)$ in $(5+1)$-dimensional CFT.  Using the $2$ and $4$-dimensional cases as motivation, we hypothesize that the $\log \epsilon$ divergence in the entanglement entropy will schematically be of the form $S_\Sigma |_{\text{log}} = (B_1 \, f_1(k) + B_2 \, f_2 (k) + B_3 \, f_3 (k) + A \, f_4 (k) ) \log \epsilon$, where the 4 functionals $f_i (k)$ are invariant under diffeomorphisms on $\Sigma$ and conformal transformations which keep $k^0_{ab} = 0$.  We find a basis of functionals for the $f_i(k)$.  In the case where the anomaly coefficients are related by $B_3 = [B_2 - (B_1/2)]/3$, we can do much better and precisely calculate the coefficient of $\log \epsilon$ in terms of the anomaly coefficients and the extrinsic curvature.  A special case of this result is found in the interacting ${\cal N} = (2,0)$ theory describing a large number $N$ of coincident M5-branes (see, for example,~\cite{Witten:1995zh,Seiberg:1997ax,Klebanov:1997kc,Harvey:1998bx,Henningson:1998gx}), for which all of the anomaly coefficients are related in a simple way.  
 
 In the case of the interacting ${\cal N} = (2,0)$ theory we use the fact that the theory is dual to the low-energy limit of M-theory on $\text{AdS}_7 \times S_4$ and utilize the holographic procedure for calculating entanglement entropy~\cite{Ryu:2006bv, Ryu:2006ef, Nishioka:2006gr, Klebanov:2007ws}.  For the more general case where the only relation between the central charges is $B_3 = [B_2 - (B_1/2)]/3$, we proceed by using the prescription in~\cite{Hung:2011xb,deBoer:2011wk} for calculating entanglement entropy holographically with Lovelock gravity~\cite{Lovelock:1971yv}.  
 
 When $\Sigma$ is allowed to have conical singularities we find new divergences in the entanglement entropy ranging in severity from $1/\epsilon^3$ to $\log^2 \epsilon$ depending on the codimension $p$ of the singular locus, with $p = 1, \dots,4$.  The most singular behavior comes from entangling surfaces for which $p = 1$.  The $\log^2 \epsilon$ divergence occurs when $\Sigma$ has a conical singularity located at a point within $\Sigma$ ($p=4$), and in this case we find that the coefficient of the $\log^2 \epsilon$ term depends simply on elementary functions of the opening angle of the cone.  Similar behavior has been found in $(1+1)$-dimensional CFT~\cite{Casini:2006hu,Drukker:1999zq,Hirata:2006jx}, where $\Sigma$ can have cusp ($p=1$) singularities, and in $(3+1)$-dimensional CFT~\cite{Klebanov:2012yf}, where $\Sigma$ can have wedge ($p=1$) and cone ($p = 2$) singularities.     
 
 A novel check of Solodukhin's formula in~\eqref{threegen} was given in~\cite{Huerta:2011qi}, where the coefficient of the $1/(mR)$ term in the IR expansion of the entanglement entropy across a circle of radius $R$ for massive free scalar and Dirac fields in $(2+1)$-dimensions was shown to be related to the coefficient of the $\log \epsilon$ term in the entanglement entropy of the corresponding massless theory in $(3+1)$-dimensions.  The entangling surface in $(3+1)$-dimensions is taken to be $\Sigma = S^1 \times S^1$, where the first circle has a large radius $L$ and the second has a radius $R \ll L$.  The author then numerically calculated the coefficient of the $1/(mR)$ term in $(2+1)$-dimensions using the methods developed in~\cite{Casini:2009sr,Srednicki:1993im,2003JPhA...36L.205P}.  Perfect agreement was found between the numerical calculation and the calculation based on the relation to $(3+1)$-dimensional entanglement entropy.  This provided a powerful check of~\eqref{threegen}.        
 
 We perform an analogous check on our formula for the coefficient of the $\log \epsilon$ term in the $(5+1)$-dimensional CFT entanglement entropy.  In this case the $1/(mR)^3$ term in the IR expansion of the massive $(2+1)$-dimensional free field entanglement entropy across a circle of radius $R$ is related to the entanglement entropy of the corresponding massless theory in $(5+1)$-dimensions~\cite{Klebanov:2012yf}.  The entangling surface in $(5+1)$-dimensions is taken to be $T^3 \times S^1$, where the $3$-torus has a large volume compared to $R^3$, and $R$ is the radius of the circle.  The constraint $B_3 = [B_2 - (B_1/2)]/3$ in our $(5+1)$-dimensional formula means that we must consider the free massive chiral multiplet in $(2+1)$-dimensions, which consists of two real scalars and a Dirac fermion.  We find good agreement between the numerical calculation in $(2+1)$-dimensions and the prediction of our $(5+1)$-dimensional formula.
 
 Additionally, we numerically calculate the renormalized entanglement entropy~\cite{Liu:2012ee} for the massive Dirac fermion in $(2+1)$-dimensions.  The renormalized entanglement entropy is defined in terms of the entanglement entropy $S(R)$ across the circle of radius $R$ by ${\cal F}(R) \equiv - S(R) + R S'(R)$.  At conformal fixed points it was shown in~\cite{Casini:2010kt, Casini:2011kv} that ${\cal F}(R) = F$, where $F$ is the finite part of the free energy of the CFT conformally coupled to the round $S^3$.  For a field theory that flows from a UV CFT to an IR CFT, it was conjectured
\cite{Myers:2010tj,Jafferis:2011zi,Casini:2011kv,Klebanov:2011gs} that $F_\text{UV} > F_\text{IR}$.  Casini and Huerta~\cite{Casini:2012ei} showed that ${\cal F}(R)$ should be a monotonic interpolating function along the RG flow between the UV and IR fixed points.  We verify that this is the case for the massive Dirac fermion.  A similar check was performed for the free massive scalar in~\cite{Liu:2012ee}.          

While this paper was in its final stages of preparation, we learned of the work of~\cite{Myers:2012ms}, which partially overlaps with our results in section~\ref{singular} on entanglement entropy across singular entangling surfaces.

\section{The shape dependence of EE in $(5+1)$-dimensional CFT}

The origin of the logarithmically divergent term in $(5+1)$-dimensional entanglement entropy is the conformal anomaly (see, for example,~\cite{Bastianelli:2000hi,Henningson:1998gx})  
\es{confanom6}{
\left< T^{\mu}\,_\mu \right> = \sum_{n=1}^3 B_n I_n + A E_6 \,,
} 
where $T_{\mu \nu} = -(2 / \sqrt{- g}) \delta I / \delta g^{\mu \nu}$ is the stress tensor, the $B_n$ and $A$ are the anomaly coefficients, $E_6$ is the $6$-dimensional Euler density, and the $I_n$ are geometric invariants constructed from the Weyl tensor $C_{\mu \nu \rho \sigma}$.  For the moment we are taking the background manifold to be a general pseudo-Riemannian manifold  ${\cal M}$.  In our conventions 
\es{I6conventions}{
I_1 &= C_{\mu \nu \rho \sigma} C^{\nu \alpha \beta \rho} C_{\alpha}\,^{\mu \sigma}\,_{\beta} \,, \qquad I_2 = C_{\mu \nu}\,^{\rho \sigma} C_{\rho \sigma}\,^{\alpha \beta} C_{\alpha \beta}\,^{\mu \nu} \,, \\
I_3 &= C_{\mu \nu \rho \sigma} \left(\nabla^2 \delta^{\mu}_\alpha +4 R^{\mu}\,_\alpha - \frac65 R \delta^{\mu}_\alpha \right) C^{\alpha \nu \rho \sigma} \,,
}
and in Euclidean signature $\int_{S^6}  \, E_6 = 2$.   

When  
there is a rotational symmetry in the space transverse to $\Sigma$, the logarithmically divergent part of the entanglement entropy can be determined~\cite{Myers:2010tj} using the replica trick 
\es{Snok6}{
S_\Sigma |_{\text{log}} &= - \int_\Sigma  \left[ 4 \pi \left(\sum_{n=1}^3 B_n { \partial I_n \over R^{\mu \nu}\,_{\rho \sigma}} g^{\perp}\,^{\mu \nu} g^{\perp}_{\rho \sigma} \right) +  A E_4  \right] \log \epsilon \,,
}
where $g^\perp_{\mu \nu} = n^{i}_\mu n^i_{\nu}$.  The rotational symmetry requirement forces us to consider entangling surfaces with vanishing extrinsic curvature.  Hung et al.~\cite{Hung:2011xb} have proposed that when the restriction of rotational symmetry in the space transverse to $\Sigma$ is lifted but one maintains $k^i_{ab} = 0$, the result in~\eqref{Snok6} is modified by a term proportional to $B_3$.  The expression in~\eqref{Snok6} is invariant under conformal transformations $g_{\mu \nu} \to e^{-2 \omega} g_{\mu \nu}$ for which $n^{i \mu} \partial_\mu \omega = 0$.  These are conformal transformations which preserve $k^i_{a b} = 0$.  The term that fails to be invariant under more general conformal transformations is the one proportional to $B_3$.    

Notice that when the background metric is conformally flat, the expression in~\eqref{Snok6} vanishes.  In this case the only contributions to the logarithmically divergent part of the entanglement entropy come from the extrinsic curvature. 

\subsection{EE with non-vanishing extrinsic curvature}      

We now calculate the dependence of the entanglement entropy on the extrinsic curvature of the $4$-dimensional entangling surface $\Sigma$.  For simplicity we take the background to be flat $\mathbb{R}^{1,5}$, and we take vanishing extrinsic curvature in the time-like direction ($k^0_{ab} = 0$).  For now we assume that the coefficient of the $A$-anomaly term remains a topological invariant when $\Sigma$ is allowed to have extrinsic curvature.  While this is a common assumption, it will be checked holographically in section~\ref{Love6}.  With this assumption we have $S_\Sigma = (B_1 \, f_1( k) + B_2 \, f_2(k) + B_3 \, f_3(k) - A \int_\Sigma \, E_4 ) \log \epsilon$, where the functionals $f_i(k)$ of the extrinsic curvature should be invariant under local diffeomorphisms on $\Sigma$ and conformal transformations which keep $k^0_{ab} = 0$.  They should also vanish when $\Sigma$ is the round $S^4$, since the CFT entanglement entropy across the round $S^4$ is equivalent  
to the Euclidean free energy of the CFT conformally coupled to $S^6$~\cite{Casini:2010kt, Casini:2011kv}.  The $\log \epsilon$ term in the $S^6$ free energy is proportional to the $A$-anomaly coefficient (see, for example,~\cite{Cardy:1988cwa}).  

We need to find 3 appropriate functionals of the extrinsic curvature to span the $f_i (k)$.
We first consider functionals which only depend on the extrinsic curvature and not on derivatives of the extrinsic curvature.  The most general form of such a functional is
\es{TFGEN}{
T &= \int_\Sigma \, \left( \alpha_1 \, k^4 +\alpha_2 \, k^2 \tr k^2 + \alpha_3 \, (\tr k^2)^2 +\alpha_4 \, k \tr k^3 + \alpha_5 \, \tr k^4 \right) \,,
}
where we are using the notation $\tr k^n = k^1_{a_1}\,^{a_2} k^1_{a_2}\,^{a_3} \cdots k^1_{a_n}\,^{a_1}$, and the coefficients $\{\alpha_1, \dots, \alpha_5 \}$ are to be determined.  Under a conformal transformation $g_{\mu \nu} \to e^{-2 \omega} g_{\mu \nu}$, the extrinsic curvature transforms non-covariantly,  
\es{exTrans}{
k_{ab}^1 = e^{- \omega} \left( k_{ab}^1 + \gamma_{ab} n^{1 \mu} \partial_\mu \omega \right) \,,
}
with $\gamma_{ab}$ the induced metric on $\Sigma$.  The non-covariant factor on the right hand side of~\eqref{exTrans} is removed by subtracting off the trace of $k_{ab}^1$.  That is, we define the traceless tensor 
\es{Kimp}{
\bar k_{ab} = k_{ab}^1 - { \gamma_{ab} \over 4} k \,,
}
so that under conformal transformations $\bar k_{ab} \to e^{- \omega} \bar k_{ab}$.  Using the improved tensor $\bar k_{ab}$ we may immediately write down the two independent, conformally invariant functionals of the extrinsic curvature:    
 \es{TF}{
T_1 &=  \int_{\Sigma} \, \left( \tr \bar k^2 \right)^2 = \int_\Sigma \, \left( (\tr k^2)^2 - \frac12 k^2 \tr k^2 + \frac{1}{16} k^4 \right) \,, \\
 T_2 &= \int_{\Sigma}  \, \tr \bar k^4 =  \int_\Sigma \, \left( \tr k^4 - k \tr k^3 + \frac{3}{8} k^2 \tr k^2 - \frac{3}{64} k^4 \right) \,. \\
}      
 
 The third functional should then involve derivatives of the extrinsic curvature.  The most general form of such a functional is
\es{jh}{
T_3 &= \int_\Sigma \, \left( (\nabla_a k)^2 + c_1 \, k^4 +c_2 \, k^2 \tr k^2 + c_3 \, (\tr k^2)^2 +c_4 \, k \tr k^3 + c_5 \, \tr k^4  \right) \,,
}   
where the constants $\{c_1, \dots, c_5\}$ are to be determined.  We will determine them later in this section, but for now we can immediately write down one constraint on the $c$'s.  This is the constraint which comes from the requirement that $T_3$ must vanish on the round $S^4$,
\es{Aconst}{
4^3 \,c_1 + 4^2 \,c_2 + 4 \,c_3 + 4 \,c_4 + \,c_5 = 0 \,.
}

An important observation is that when the background manifold ${\cal M}$ is allowed to have a nonzero curvature $R_{\mu \nu \rho \sigma}$, the functionals $T_1$ and $T_2$ remain conformally invariant functionals without the need for adding any new terms proportional to the background curvature.   
This cannot be the case for the functional $T_3$.  This can be seen quite simply be restricting to conformal transformations $g_{\mu \nu} \to e^{-2 \omega} g_{\mu \nu}$ such that $n^{i\mu} \partial_\mu \omega = 0$.  In this case $k \to e^{+\omega} k$ acts just like a scalar field in 4-dimensions.  It is well known that the action of a scalar field conformally coupled to curvature on a 4-dimensional manifold $\Sigma$  is 
\es{scalaraction}{
I = {1 \over 2} \int_\Sigma \left( (\nabla_a k)^2 + \frac{{\cal R}}{6} k^2 \right) \,,
}     
where ${\cal R}$ is the curvature scalar on $\Sigma$.  The curvature scalar ${\cal R}$ is related to the curvature of ${\cal M}$ by the contracted Gauss-Codazzi equation, ${\cal R} = R^{ab}\,_{ab} + k^2 - \tr k^2$, where $R_{abcd}$ denotes the projection of $R_{\mu \nu \rho \sigma}$ onto $\Sigma$.  The action $I$ has a term proportional to the integral over $R^{ab}\,_{ab}$ time $k^2$.  We can similarly expect $T_3$ to be modified by terms proportional to the background curvature when the background curvature is non-vanishing.    

Recall that~\eqref{Snok6}, which gives the entanglement entropy in the case where there is a rotational symmetry in the space transverse to $\Sigma$, is only invariant under conformal transformations for which $n^{i \mu} \partial_\mu \omega = 0$.  The term that fails to be invariant is proportional to the anomaly coefficient $B_3$.  This leads us to conjecture that in the opposite limit, when $k^1_{ab}$ is non-zero but $R_{\mu \nu \rho \sigma} = 0$, the functional $T_3$ will enter the entanglement entropy through a term proportional to $B_3$.  This conjecture will be checked holographically later in this section.  

To summarize the above discussions, we have narrowed the form of the $\log \epsilon$ term in the entanglement entropy to 
\es{EE6log1}{
S_\Sigma |_{\text{log}} = \left[ \left(\sum_{i=1}^3 \mu^{(1)}_i B_i \right) T_1 + \left(\sum_{i=1}^3 \mu^{(2)}_i B_i \right) T_2 + \mu^{(3)} \, B_3 \, T_3 - A \int_\Sigma \, E_4 \right] \log \epsilon \,,
} 
where the $7$ coefficients $\{ \mu^{(1)}_i, \mu^{(2)}_i, \mu^{(3)}\}$ have yet to be determined.  When the anomaly coefficients satisfy certain relations which allow us to use Einstein or Lovelock gravity in the bulk, we can precisely determine these coefficients through holographic calculations.

\subsubsection{EE in the interacting ${\cal N} = (2,0)$ theory}

In the case of the interacting ${\cal N} = (2,0)$ theory describing a large number $N$ of coincident M5-brane, it is convenient to write~\eqref{EE6log1} in the form 
\es{EE20GEN}{
S_\Sigma |_{\text{log}} = -N^3 \left[ {2 \over 3} \int_\Sigma \sqrt{\gamma}\, E_4 + { \mu^{(1)} T_1 + \mu^{(2)} T_2 + \mu^{(3)} T_3 \over   96 \pi^2}  \right] \log \epsilon \,,
} 
where we have used the anomaly coefficients~\cite{Henningson:1998gx,Bastianelli:2000hi}\footnote{Note that our $A$ anomaly coefficient differs from those in~\cite{Henningson:1998gx,Bastianelli:2000hi} and also the recent work of~\cite{Elvang:2012st} because we normalize the Euler density $E_6$ so that $\int_{S^6} \, E_6 = 2$, while those references choose a different normalization.  For example, in~\cite{Bastianelli:2000hi} they use the normalization $-{1 \over 8 \cdot 3!} {1 \over (4 \pi)^3 } \int_{S^6} \, E_6 = 2$, and in~\cite{Elvang:2012st} they use the normalization ${1 \over \cdot 3!} {1 \over (4 \pi)^3 } \int_{S^6} \, E_6 = 2$.}
\es{20anomCoeff}{
A  &= \frac{8 \cdot 3!}{7!} \left({35 \over 2}\right) (4 N^3)  \,, \qquad B_1 = \frac{- 1}{(4 \pi)^37!} \left( 1680  \right) (4 N^3) \,, \\
    B_2 &= \frac{-1}{(4 \pi)^37!} \left( 420 \right) (4 N^3) \,, \qquad B_3 = \frac{1}{(4 \pi)^37!} \left( 140 \right) (4 N^3) \,. 
}

We may calculate the entanglement entropy holographically by finding the $5$-dimensional surface $\Sigma_5$, which approaches the entangling surface $\Sigma$ at the boundary of $AdS_7$, is extended in the rest of the spatial dimensions, and minimizes the area functional
\es{areaFunctional6}{
S_\Sigma = {1 \over 4 G_N^{(7)}} \int_{\Sigma_5} d^5\sigma \sqrt{\det G_{{\text{ind}}}^{(5)}} \,,
}
where $G_{\text{ind}}^{(5)}$ is the induced metric on $\Sigma_5$.  We write the $AdS_7$ metric as
\es{ads7g}{
ds_7^2 = L_{AdS}^2 {dy^2 + dx_{\mu} dx^{\mu} \over y^2}  \,, \qquad y \geq 0 \,,
} 
with $dx_{\mu} dx^{\mu}$ the metric on $\mathbb{R}^{1,5}$, $L_{AdS} = 2 \ell_p (\pi N)^{1/3}$, $G_N^{(7)} = 6 \, G_N^{(11)} / (\pi^2 L_{AdS}^4)$, and $G_N^{(11)} = 16 \pi^7 \ell_p^9$.  

We consider entangling surfaces of the form $\Sigma = T^{4-n} \times \tilde \Sigma_n$, where each $\tilde \Sigma_n$ is a deformed $n$-sphere and $1 \leq n \leq 4$.  In the case $n=4$ we simply have $\Sigma = \tilde \Sigma_4$.  We will take the volume of each torus $T^{4-n}$ to be $L^{4-n}$, with $L$ large compared to other scales.  In each case we write the metric on $\mathbb{R}^{1,5}$ as 
\es{metric15}{
dx_{\mu} dx^{\mu} = -dt^2 + \sum_{i=1}^{4-n} dx_i^2+ dr^2 + r^2(d \theta^2 + \sin^2\theta \, d \Omega_{n-1}^2 ) \,,
}
where $d \Omega_{n-1}^2$ is the metric on the unit $S^{n-1}$.  We write the spacetime metric this way because we will take $\tilde \Sigma_n$ to wrap the $(n-1)$-sphere and be given by the curve $r = R(\theta)$.  The function $R(\theta)$ must be smooth, positive and have vanishing derivative at its endpoints.  The latter condition ensures that our entangling surface is free of conical singularities.  The torus $T^{4-n}$ is extended in the $x_i$ directions, which are compactified  such that $x_i \sim x_i + L$.  We choose such entangling surfaces because they are not overly complicated to describe but still provide non-trivial geometries.

Before beginning the calculations, note that it is actually sufficient to just consider $\Sigma = T^{2} \times \tilde \Sigma_2$ or $\Sigma = T^{1} \times \tilde \Sigma_3$ alone.  That is, each one of these entangling surfaces is general enough to determine the undetermined parameters.  We provide an analysis of all four entangling surfaces, however, to show that the method is consistent.  In addition to the calculations presented below, we have also performed numerical calculations where the radial coordinate $r$ in $\tilde \Sigma_n$ is taken to be a function of $\theta$ and the $n-1$ coordinates in the internal $S^{n-1}$.  Those calculations were consistent with the general findings below.     

\subsubsection*{{\bf $\Sigma =T^3 \times \tilde \Sigma_1$}}

When $\Sigma = T^3 \times \tilde \Sigma_1$ the functionals $T_1$, $T_2$~\eqref{TF} and $T_3$~\eqref{jh} reduce to 
\es{reducedinvarfunctionals}{
T_1 &=   {9 \, L^3 \over 16}  \int_{\tilde \Sigma_1} \, \kappa^4 \,, \qquad T_2 =   {21 \, L^3 \over 64}  \int_{\Sigma_1} \, \kappa^4 \,, \\
 T_3 &= L^3 \int_{\tilde \Sigma_1} \, \left[ (\nabla \kappa)^2 + c_1 \, \kappa^4 +c_2 \, \kappa^2 \tr k^2 + c_3 \, (\tr \kappa^2)^2 +c_4 \, \kappa \tr \kappa^3 + c_5 \, \tr \kappa^4  \right] \,,
}
where 
\es{kappa}{
\kappa = {R^2(\theta) + 2 R'^2(\theta) - R(\theta) R''(\theta) \over \left( R^2(\theta) + R'^2(\theta) \right)^{3/2} }
}
 is the extrinsic curvature of $\tilde \Sigma_1$ in $\mathbb{R}^2$.  We take the bulk entangling surface to be specified by the function $r = r(y,\theta)$.  Near the boundary at $y = 0$ the solution to the equation of motion is 
\es{r7soln}{
 r(y,\theta) = R(\theta) - {\kappa(\theta) \sqrt{R(\theta)^2 + R'(\theta)^2} \over 8 R(\theta)} y^2 + O(y^4) \,.
 }
 Substituting the solution into~\eqref{areaFunctional6} we calculate the logarithmically divergent term in the entanglement entropy, 
 \es{Slog7}{
 S_\Sigma |_{\text{log}} = -\frac{N^3}{96 \,\pi^2}\int_{\tilde \Sigma_1} \, \left( (\nabla k)^2 - {9 \kappa^4 \over 16}  \right) \log \epsilon \,.
 } 
 Comparing with~\eqref{jh} and~\eqref{EE20GEN} we find the following constraints among the undetermined coefficients,
 \es{undS1}{
 \mu^{(1)} = 1-{7 \mu^{(2)}\over 12} + {16 \over 27}\big(3 (c_1 + c_2) + 5 c_3 \big) \,, \qquad \mu^{(3)} = -1 \,.
 }
 In writing down the above expression we have made the choices  $c_4 = (8/3)\, c_3$ and $c_5 = -2 \,c_3$ for later convenience.  We are allowed to do this because we can always redefine $T_3$ by adding a linear combination of $T_1$ and $T_2$.  We will make these choices for $c_4$ and $c_5$ throughout the rest of this section.   
 
 \subsubsection*{{\bf $\Sigma =T^2 \times \tilde \Sigma_2$}}
 
 When $\Sigma =T^2 \times \tilde \Sigma_2$, the extrinsic curvature tensor on $\tilde \Sigma_2$ becomes 
 \es{S2curvT}{
 k_{ab} = {1 \over \sqrt{R^2 + R'^2 } }
 \left( \begin{array}{cc}
R^2 + 2 rR'^2 - R R'' & 0 \\
0 & R \sin^2\theta ( R - \cot\theta  R' )  \end{array} \right) \,,
 }
 from which we may calculate the functionals $T_1$, $T_2$~\eqref{TF} and $T_3$~\eqref{jh}.  The holographic calculation is analogous to the previous one.  On the bulk entangling surface we take $r = r(y,\theta)$.  The solution to the equation of motion for the minimal area surface near the boundary is
 \es{r2soln}{
 r(y,\theta) = R(\theta) + {1 \over 8 R^2} \left( \cot \theta R' - { R \,\big( 3 \, R'^2 + R(2 \, R -R'') \big) \over R^2 + R'^2 } \right) y^2 + O(y^4) \,,
 }
which allows us to calculate the logarithmically divergent term in the entanglement entropy as a function of $R(\theta)$.  Comparing with~\eqref{jh} and~\eqref{EE20GEN} we deduce the following relationships among the undetermined coefficients 
 \es{undS2}{
 c_1= {7 \over 16} + { c_3 \over 3} \,, \qquad c_2 = -1 - 2\, c_3  \,, \qquad  \mu^{(1)} = \mu^{(2)} = 0 \,, \qquad \mu^{(3)} = -1 \,.  
 }  

 \subsubsection*{{\bf $\Sigma =S^1 \times \tilde \Sigma_3$}}
 
 With $\Sigma = S^1 \times \tilde \Sigma_3$ the extrinsic curvature tensor entries $k_{ab}$ on $\tilde \Sigma_3$ are the same as those in~\eqref{S2curvT} for $a,b = 1, 2$.  The new non-zero entry is  
  \es{S3curvT}{
 k_{33}= { \sin^2 \theta_2 \sin^2 \theta (R - \cot \theta R' )  \over \sqrt{R^2 + R'^2 } }  \,,
 }
 where, using the notation of~\eqref{metric15}, we write $d \Omega_2^2 = d \theta_2^2 + \sin^2\theta_2 d \phi^2$.  The procedure described above for calculating the entanglement entropy holographically may be carried out in an analogous fashion.  The solution for $r(y,\theta)$ near the boundary is 
 \es{r3soln}{
 r(y,\theta) = R(\theta) + {1 \over 8 R^2} \left( 2 \cot \theta R' - { R \,\big( 4 \, R'^2 + R(3 \, R -R'') \big) \over R^2 + R'^2 } \right) y^2 + O(y^4) \,.
 }
 Calculating the logarithmically divergent term in the entanglement entropy and comparing with~\eqref{jh} and~\eqref{EE20GEN}, we find the same set of equations as in~\eqref{undS2}. 

\subsubsection*{{\bf $\Sigma =  \tilde \Sigma_4$}}

The final case we will consider is when $\Sigma = \tilde \Sigma_4$.  The extrinsic curvature tensor entries $k_{ab}$ on $\Sigma_4$ are the same as on $\tilde \Sigma_3$, which was discussed in the previous paragraph, for $a,b = 1,2,3$.  The new non-zero entry is 
 \es{S4curvT}{
 k_{44}= { \sin^2 \theta_3 \sin^2 \theta_2 \sin^2 \theta (R - \cot \theta R' )  \over \sqrt{R^2 + R'^2 } }  \,,
 }
 where we write $d \Omega_3^2 = d \theta_3^2 + \sin^2\theta_3 d \Omega_2^2$.  In the holographic calculation we find 
 \es{r4soln}{
 r(y,\theta) = R(\theta) + {1 \over 8 R^2} \left( 3 \cot \theta R' - { R \,\big( 5 \, R'^2 + R(4 \, R -R'') \big) \over R^2 + R'^2 } \right) y^2 + O(y^4) 
 }
 near the boundary at $y = 0$.  Calculating the entanglement entropy we deduce that
  \es{undS4}{
 c_1= {7 \over 16} + { c_3 \over 3} \,, \qquad c_2 = -1 - 2\, c_3  \,, \qquad  \mu^{(1)} = - {7 \, \mu^{(2)} \over 12} \,, \qquad \mu^{(3)} = -1 \,.  
 }

The fact that $c_3$ is not determined in the above analysis reflects the fact that in this theory $E_4$ can mix with $T_3$, since $B_3$ and $A$ are related.  However, with the constraint in~\eqref{Aconst} we are led to take $c_3 = -25/16$.  With this choice $T_3$ becomes equal to 
  \es{T3invarfunctionals}{
T_3 &= \int_\Sigma \, \left( (\nabla_a k)^2 - {25 \over 16} k^4 +11\, k^2 \tr k^2 - 6 (\tr k^2)^2 - 16 \, k \tr k^3 + 12\, \tr k^4  \right) \,,
}
and we arrive at our final formula for the shape dependence of entanglement entropy in the interacting ${\cal N} = (2,0)$ theory:
 \es{Slog20}{
  S_\Sigma |_{\text{log}} = -N^3 \left[ {2 \over 3} \int_\Sigma \, E_4 + {T_3 \over 96 \pi^2}  \right]\log \epsilon \,.
 }

\subsubsection{EE using Lovelock gravity}  \label{Love6}

In this section we generalize~\eqref{Slog20} to theories with three independent anomaly coefficients.  
The single constraint on the anomaly coefficients is $B_3 = [B_2 - (B_1/2)]/3$, which allows us to 
compute the entanglement entropy holographically using Lovelock gravity.  
In Lovelock gravity the entanglement entropy functional in~\eqref{areaFunctional6} is modified to~\cite{Hung:2011xb}
\es{areaFunctional6LV}{
S_\Sigma = \frac{1}{4 G_N^{(7)}} \int_{\Sigma_5} d^5\sigma \sqrt{G_{{\text{ind}}}^{(5)}}\left[1 + \frac\lambda6 f_{\infty} L_{AdS}^2 \, {\cal R}_5 - \frac\mu8 f_{\infty}^2 L_{AdS}^4 \, ({\cal R}_5\,_{\mu \nu \rho \sigma} {\cal R}_5^{\mu\nu\rho\sigma} - 4 {\cal R}_5\,_{\mu \nu} {\cal R}_5^{\mu \nu} + {\cal R}_5^2) \right] \,,
}
where $ {\cal R}_5\,_{\mu \nu \rho \sigma}$ is the curvature tensor of the induced metric on the holographic entangling surface $\Sigma_5$.  We leave out a Gibbons-Hawking boundary term in~\eqref{areaFunctional6LV} since it will not be involved in our discussion.  The parameter $f_\infty$ is a positive root of the equation 
  \es{finfty6}{
1 = f_\infty - f_\infty^2 \lambda - f_\infty^3 \mu \,.
}
The gravitational theory has an AdS vacuum with radius $L_{AdS}$.  The central charges of the dual CFT are given by~\cite{deBoer:2009gx}
\es{centralcharges6}{
B_1 &= \frac{L_{AdS}^5}{8 \pi G_N^{(7)}} \frac{-9 + 26 f_\infty \lambda + 51 f_\infty^2 \mu}{288} \,, \qquad B_2 = \frac{ L_{AdS}^5}{8 \pi G_N^{(7)}} \frac{-9 + 34 f_\infty \lambda + 75 f_\infty^2 \mu}{1152} \,, \\
B_3 &= \frac13 \left(B_2 - {1 \over 2} B_1\right)   \,, \qquad A = \pi^3  \frac{L_{AdS}^5}{8 \pi G_N^{(7)}}  \frac{3 -10 f_\infty \lambda -45 f_\infty^2 \mu}{3} \,.
} 

We may calculate the entanglement entropy in an analogous fashion to the way described in the previous subsection for the set of entangling surfaces $\Sigma = T^{4-n} \times \tilde \Sigma_n$ given in that section.  The key insight is that, at least in the examples we use, the solutions to the equation of motion do not change near the boundary in the presence of nonzero $\lambda$ and $\mu$.  What is required, then, is to take the functions $r = r(y,\theta)$ given in~\eqref{r7soln},~\eqref{r2soln},~\eqref{r3soln},~\eqref{r4soln} for $n = 1, ..., 4$, respectively, and use them in~\eqref{areaFunctional6LV} to evaluate the logarithmically divergent term in the entanglement entropy.  Carrying out this procedure leads uniquely to
   \es{Slog7Fin2}{
 S_\Sigma|_{\text{log}} &=  - \left( A \int_\Sigma \, E_4 + 6 \, \pi \left[3  \left(B_2 - {B_1 \over 4} \right) J + B_3 \, T_3 \right] \right) \log \epsilon \,,
 } 
 where 
  \es{Jinvarfunctionals}{
J =  \int_\Sigma \, \left[ {5 \, k^2 \over 4} \left( {k^2 \over 8} - \tr k^2 \right) + (\tr k^2)^2 + 2(k \tr k^3 - \tr k^4) \right]  \\
}  
and $T_3$ is given in~\eqref{T3invarfunctionals}.  We should emphasize that~\eqref{Slog7Fin2} is only valid for theories which satisfy the constraint $B_3 = [B_2 - (B_1/2)]/3$.  With this constraint there are three independent anomaly coefficients.  This is the reason why there are only three independent functionals of the extrinsic curvature in~\eqref{Slog7Fin2}, while in a generic theory we would expect four (see~\eqref{EE6log1}).       

A simple, relevant example of the above procedure is when $\Sigma = T^3 \times S^1$, with the $S^1$ of constant radius $R$.  In this case the solution for $r(y)$ in~\eqref{r7soln} near the boundary reduces to $r(y) = R - (y^2/8R) + O(y^4)$.  Evaluating the entanglement entropy functional in~\eqref{areaFunctional6LV} on this solution, we find that
\es{S31ExampleLV}{
S_\Sigma = {L_{\text{AdS}}^5 \, L^3 \, \pi \over 2 G_N^{(7)} } \int_{\epsilon} dy \, &\left[ {R \, \big( 3 - 5 \, f_{\infty} (2 \, \lambda + 9 f_{\infty} \mu)\big) \over 3 y^5} - {3 (1 - 2 f_{\infty} \lambda - 3 f_\infty^2 \mu) \over 32 \, R\, y^3} \right. \\
 &\left.- {3 \big( 3 - 2 f_\infty \lambda + 3 f_\infty^2 \mu \big) \over 2048 \, R^3 \,y} + O(y^0) \right] \,,
}  
which leads to 
 \es{Slog7Fin}{
 S_\Sigma|_{\text{log}} = {9 \, L^3 \, \pi^2 (20 \, B_2 - 7 \, B_1 ) \over 32 \, R^3 } \log \epsilon \,.
 }
  
 \subsection{EE with conical singularities} \label{singular}
 
 It is interesting to ask what new types of singularities the entanglement entropy can have when the entangling surface $\Sigma$ in $(5+1)$-dimensional CFT is allowed to have conical singularities.  The conically singular entangling surfaces can be categorized into four different types depending on the codimension $p$ of the singular locus, with $p = 1, \dots, 4$. 
We will consider conically singular entangling surfaces of the form $\Sigma =T^{(4-p)} \times {\cal C}^{(p)}$, where $T^{(4-p)}$ is the $(4-p)$-dimensional torus and ${\cal C}^{(p)}$ is the $p$-dimensional cone of opening angle $\Omega$.

The singular entangling surfaces are taken to lie in the $t=0$ hyperplane of flat Minkowski space.  For each $p$ we write the spacetime metric as $ds^2 = -dt^2 +\sum_{i=1}^{4-p} dx_i^2 + dz^2 + dr^2 + r^2 d \Omega_{p-1}^2$, with $-\infty < z < \infty$ and $0 \leq r < \infty$.  The torus $T^{(4-p)}$ wraps the directions $x_i$, which are compactified such that $x_i \sim x_i + L$ with $L$ large compared to other scales.  The cone ${\cal C}^{(p)}$ is taken to wrap the unit $S^{p-1}$ and be given by the line $z = (\cot \Omega) \, r$.

Even though~\eqref{Slog7Fin2} was derived under the assumption of smooth $\Sigma$, we can still try and use this equation to figure out the form of the new singularities in the entanglement entropy in the presence of conical singularities.  This method was applied successfully in $(3+1)$-dimensional CFT in~\cite{Klebanov:2012yf}.  Roughy speaking the integrand in~\eqref{Slog7Fin2} goes as $1/r^4$ while the integration measure brings in a factor of $dr \, r^{p-1}$ for $p > 1$.  The case $p = 1$ is an exception since the extrinsic curvature of ${\cal C}^{(1)}$ vanishes everywhere except at the singular locus, where it diverges.  Restricting to $p > 1$ we can do the integral over $r$ with the short distance cut-off $r > \epsilon$, and we find that when $p < 4$ we get divergences in the entanglement entropy of the form $f(\Omega) L^{4-p} (\log \epsilon) / \epsilon^{4-p}$, with $f(\Omega)$ some function of the opening angle, plus less severe divergences.  
By arguments directly analogous to those presented in~\cite{Klebanov:2012yf}, one can show that when $p=1$ the most severe, new divergence the entanglement entropy gets is an $f(\Omega) L^3 / \epsilon^3$ type divergence.  A simple way of heuristically arriving at this result is to regulate the singular tip of ${\cal C}^{(1)}$ by a semicircle times a $3$-torus.     

We will be more precise in the case $p = 4$.  Writing the metric on the $S^3$ as $d \Omega_3^2 = d \theta_1^2 + \sin^2\theta_1 ( d \theta_2^2 + \sin^2 {\theta_2} \,d \phi^2 )$, we find that the nonzero components of the extrinsic curvature are
\es{knonzero}{
k_{\theta_1 \theta_1} = r \cos \Omega \,, \qquad k_{\theta_2 \theta_2} =k_{\theta_1 \theta_1} \sin^2 \theta_1 \,, \qquad k_{\phi \phi} =k_{\theta_2 \theta_2} \sin^2 \theta_2 \,.
} 
Naively using~\eqref{Slog7Fin2} then gives the $\log^2 \epsilon$ divergence 
   \es{Slog7SING}{
 S_{{\cal C}^{(4)}} |_{\log^2 \epsilon}=  { 9 \, \pi^3 \over 128} {\cos^2 \Omega \over \sin \Omega} \left[-121 \, B_1 + 236 \, B_2 + (7 \, B_1 - 20 \, B_2) \cos(2\, \Omega) \right] \log^2 \epsilon   \,
 }
 in theories where $B_3 = [B_2 - (B_1/2)]/3$.  A similar $\log^2 \epsilon$ divergence was found in $(3+1)$-dimensional CFT in~\cite{Klebanov:2012yf} with conical entangling surfaces.  As in that case, the function in~\eqref{Slog7SING} which multiplies $\log^2 \epsilon$ diverges like $1/ \Omega$ as $\Omega$ goes to zero, and it approaches zero as $\Omega$ goes to $\pi/2$, where the conical singularity disappears.  We find that the first correction away from $\Omega = \pi/2$ comes from the functional $T_3$ and is thus proportional to $B_3$: 
 \es{Slog6SINGB3}{
 S_{{\cal C}^{(4)}} |_{\log^2 \epsilon}=  \left[54 \, \pi^3 \, B_3 \left(\Omega - {\pi \over 2} \right)^2 + O\left(\Omega - {\pi \over 2} \right)^4 \right] \log^2 \epsilon \,.  
 }     
 
 At small $\Omega$~\eqref{Slog7SING} becomes  
 \es{smallOmega}{
  S_{{\cal C}^{(4)}} |_{\log^2 \epsilon} = - \left( {27\, \pi^3 ( 19 \, B_1 - 36 \, B_2 ) \over 64 \, \Omega} + O(\Omega) \right) \log^2 \epsilon \,.
 }
 We may understand this result in a simple way because at small $\Omega$ we may approximately decompose the cone ${\cal C}^{(4)}$ into the union of ``cylinder-like'' entangling surfaces of the form $I \times S^3$, where $I$ denotes the interval of infinitesimal length $d L$, and the $S^3$ has a radius $R = L \sin \Omega$.  Here $L$ denotes the distance between the apex of the cone and the ``cylinder-like'' surface.  Using~\eqref{Slog7Fin2} we determine that the entanglement entropy across this surface is 
 \es{SS1S3SING}{
 S_{\Sigma} |_{\text{log}} = {27 \, \pi^3 (19\, B_1 - 36\, B_2) \over 32 \, \sin \Omega} { d L \over L} \log( \epsilon / L ) \,. 
 } 
Integrating the above expression over $L$ and taking $\Omega$ to be small gives exactly the expression in~\eqref{smallOmega}.      

 There is a small subtlety in deriving~\eqref{Slog7SING} which also arises in the $(3+1)$-dimensional case~\cite{Klebanov:2012yf}.  Since we cannot take the logarithm of a dimensionfull quantity, we know that the $\log \epsilon$ term in~\eqref{Slog7Fin2} must be accompanied by a finite term proportional to $\log r$ such that we can write $\log \epsilon \to \log (\epsilon / r)$.  The finite term must be proportional to $ \log r$ since the radius $r$ is the only other quantity with dimensions of length in the problem.  In deriving~\eqref{Slog7SING} one is then led to perform the integral $\int_\epsilon {dr \over r} \, { \log ( \epsilon / r)} = -{1 \over 2} \log^2 \epsilon + O(\log \epsilon)$.  This differs by a factor of $1/2$ from the naive expectation $\int_\epsilon {dr \over r} \, { \log ( \epsilon)} = - \log^2 \epsilon + O(\log \epsilon)$. 
 
 We now give a more precise, holographic derivation of~\eqref{Slog7SING} along the lines of that presented in~\cite{Klebanov:2012yf}.  This derivation will lead to the same factor of $1/2$ that we motivated heuristically in the previous paragraph.   We write the $AdS_7$ metric as 
 \es{AdS7metricSING}{
 ds^2 = {dy^2 - dt^2 + dr^2 + r^2 \, ( d \theta^2 + \sin^2 \theta \, d \Omega_3^2 ) \over y^2} \,,
 } 
 with the boundary entangling surface specified by $\theta = \Omega$ at $t = 0$.  When the bulk is described by Einstein gravity ($\lambda = \mu = 0$) the entanglement entropy functional becomes 
 \es{EINSING}{
 S_{{\cal C}^{(4)}} = {\pi^2 \over 2 G_{N}^{(7)} } \int dr \, r^2 \int d \theta {\sin^3 \theta \over y^5(r,\theta)} \sqrt{ (\partial_\theta y)^2 + r^2 \big( 1 + (\partial_r y)^2 \big) } \,,
 }
 where we have taken $y = y(r,\theta)$.  
   The symmetries of $AdS$ spacetime and the cone ${\cal C}^{(4)}$ allow us to take the ansatz $y(r,\theta) = r /  \tilde g(\theta)$.  It is convenient to further change variables to $s \equiv \cos \theta$, with $g(s) = \tilde g(\theta)$, and to consider $s$ as a function of $g$ instead of $g(s)$. Substituting the ansatz into~\eqref{EINSING} and making the changes of variables, we find that the entanglement entropy functional becomes the following functional for $s(g)$: 
   \es{EINSING2}{
    S_{{\cal C}^{(4)}} = {\pi^2 \over 2 G_{N}^{(7)} }  \int {dr \over r} \int dg \, g^3 \,\big(1 - s(g)^2 \big) \sqrt{1 - s(g)^2 + g^2(1 + g^2) s'(g)^2 } \,.
   } 
   
   To find the solution for $s(g)$ near the boundary we need to solve the Euler-Lagrange equation at large $g$.  A direct calculation gives 
   \es{solnSING}{
   s(g) = s_0 + {3 \, s_0 \over 8 \, g^2} + \cdots \,,
   }
   with $s_0 \equiv \cos \Omega$.  Substituting this solution into~\eqref{EINSING2} allows us to determine the $\log^2 \epsilon$ divergence in the entanglement entropy.  This term comes from performing the following integrals over $r$ and $g$:
   \es{intexample}{
   \int_{\epsilon} {dr \over r} \int^{r/\epsilon}_{g_0} dg \, {1 \over g} =  {\log^2 \epsilon \over 2} + O( \log \epsilon) \,. 
   }
   The lower bound on the $r$ integral and the upper bound on the $g$ integral are determined by the UV cut-off $\epsilon$, while $g_0$ denotes the minimum value of $g$, whose precise value is not needed to calculate the $\log^2 \epsilon$ divergence.  Using~\eqref{intexample} we find that  
   \es{EINSING3}{
   S_{{\cal C}^{(4)}} |_{\log^2 \epsilon} = {9 \, \pi^2 \over 16384 \,  G_N^{(7)} } {\cos^2 \Omega \over \sin \Omega} \left[31 - \cos(2 \Omega) \right] \log^2 \epsilon \,,
   } 
   which agrees with~\eqref{Slog7SING} when $\lambda = \mu = 0$.
   
   When $\lambda$ and $\mu$ are non-zero, we may proceed by performing a calculation analogous to the one presented above but using the full entanglement entropy functional in~\eqref{areaFunctional6LV}.  A key insight is that the solution for $s(g)$ near the boundary is still given by~\eqref{solnSING} at nonzero $\lambda$ and $\mu$.  What remains, then, is to evaluate the more complicated functional on this solution.  Doing so leads exactly to the result in~\eqref{Slog7SING}.

\section{EE in $(2+1)$-dimensional free massive theories}  \label{massiveFree}

It is a non-trivial and remarkable fact that we may check~\eqref{Slog7Fin2} by comparing with the entanglement entropy of the $(2+1)$-dimensional free massive chiral multiplet.  The relationship between the entanglement entropy of free $(2n+4)$-dimensional CFT, with integer $n \geq 0$, and that of free massive $(2+1)$-dimensional entanglement entropy is outlined in~\cite{Klebanov:2012yf} based on the work of~\cite{2005JSMTE..12..012C,Casini:2009sr,Huerta:2011qi}.

For $(2+1)$-dimensional theories theories with a mass gap of order $m$, the general structure of entanglement entropy across a smooth entangling surface $\Sigma$ is expected to have the form\cite{Casini:2005zv,Casini:2009sr,Grover:2011fa,Huerta:2011qi,Klebanov:2012yf}
\es{IRgeneral}{
	S_\Sigma = \alpha \frac{\ell_{{\Sigma}}}{\epsilon}  + \beta\, m\, \ell_{\Sigma} -\gamma +  \sum_{n=0}^{\infty}
	 \frac{\tilde c_{-1 - 2n}^{\Sigma} }{m^{2n +1} }\,,
}
where the coefficients $\tilde c^\Sigma_{-1-2n}$ are integrals of functions of the extrinsic curvature and its
derivatives \cite{Grover:2011fa,Klebanov:2012yf}, and $\gamma$ is the topological entanglement entropy \cite{Kitaev:2005dm,Levin:2006zz}.  Throughout the rest of this section we will consider the free massive theory consisting of $n_0$ real massive scalars and $n_{1/2}$ massive Dirac fermions, for which $\gamma = 0$ and $\beta = -(n_0 + n_{1/2})/12 $ (see, for example,~\cite{Solodukhin:2011gn,Fursaev:1994in,Fursaev:1995ef}).  

The coefficients $\tilde c_{-1 - 2n}^{\Sigma} $ are related to the entanglement entropy of the free massless theory in $(2 n+4)$-dimensions.      
If the entropy of the $(2 n+4)$-dimensional theory across the entangling surface $\Sigma_{2n+2} = T^{2n+1} \times \Sigma$ has the anomaly term
\es{anomTerm}{
\left. S_{ {\Sigma_{2 n + 2}}}^{(2 n + 4)} \right|_{\text{log}} = s^{(2 n + 4)}_{\Sigma_{2 n + 2}}\log( \epsilon) \,,
}
then  we can immediately read off the coefficient $\tilde c_{-1 - 2n}^{\Sigma}$ in~\eqref{IRgeneral}~\cite{Klebanov:2012yf}:
\es{acIdent}{
\tilde c_{-1 - 2n}^{\Sigma} = - \frac{ \pi (2  \pi)^n (2n-1)!! }{\Vol(T^{2n+1})} s^{(2n+4)}_{\Sigma_{2 n + 2}}\,.
}
The above formula is slightly modified for fermions.  Dirac fermions in $(2n+4)$ dimensions are in a $2^{n+2}$-dimensional representation, which after dimensional reduction reduces to $2^{n+1}$ $(2+1)$-dimensional Dirac fermions.  Thus, the right hand side of \eqref{acIdent} should be divided by $2^{n+1}$ for Dirac fermions.  For example, we can use the expression for $s_{\Sigma_2}^{(3+1)}$ in~\eqref{threegen} to write~\cite{Huerta:2011qi,Klebanov:2012yf}
\es{a2Rulekappa}{
\tilde c^{\Sigma}_{-1} = -\frac{1}{2^5 \cdot 15} (n_0 + 3 n_{1/2} ) \oint ds\, \kappa^2 \,.
}

To use our expression for $s_{\Sigma_4}^{(5+1)}$ in~\eqref{Slog7Fin2}, we need the free field anomaly coefficients in $(5+1)$-dimensions~\cite{Bastianelli:2000hi}:
 \es{freeFieldAnom}{
 A &= \frac{8 \cdot 3!}{7!} \left( \frac5{72} n^{(6)}_0 + \frac{191}{72} n^{(6)}_{1/2}   \right) \,, \quad B_1 = \frac{- 1}{(4 \pi)^37!} \left( \frac{28}{3} n^{(6)}_0 + \frac{896}{3} n^{(6)}_{1/2}  \right) \,, \\
  B_2 &= \frac{1}{(4 \pi)^37!} \left( \frac{5}{3} n^{(6)}_0 - 32 n^{(6)}_{1/2}   \right) \,, \quad B_3 = \frac{1}{(4 \pi)^37!} \left( 2 n^{(6)}_0 + 40 n^{(6)}_{1/2}  \right) \,,
 }
 where $n^{(6)}_0$ is the number of scalars and $n^{(6)}_{1/2}$ is the number of Dirac fermions.  The minimal theory which obeys the constraint $B_3 = [B_2 - (B_1/2)]/3$ is that of the free ${\cal N} = (1,0)$ hypermultiplet, consisting of one Weyl fermion and 4 real scalars.  Under dimensional reduction to $(2+1)$-dimensions this theory reduces to that of 2 free chiral multiplets.  Taking $n_0 = 2 \, n_{1/2}$ in $(2+1)$-dimensions, we find  
\es{a3Rulekappa}{
\tilde c^{\Sigma_1}_{-3} = -\frac{n_{1/2} }{2^{11}} \oint ds\, \kappa^4 + {n_{1/2} \over 2^7 \cdot 15} \oint ds \, \left( {d \kappa \over ds} \right)^2 \,.
}  
This allows us to write down the IR expansion of the renormalized entanglement entropy for a free massive chiral multiplet, 
\es{FIRchiral}{
{\cal F}(m R) = {\pi \over 24} \left(  {1 \over m R} + {3 \over 32} {1 \over (m R)^3 } + O(1/(m R)^5) \right) \,.
}
In the following subsection we check the expansion above against a numerical calculation of the renormalized entanglement entropy.

\subsection{Numerical computation of the renormalized EE}

We compute the $(2+1)$-dimensional entanglement entropy numerically for the free massive scalar and Dirac fermion following the prescription in~\cite{Huerta:2011qi}, which is based on the works of~\cite{Casini:2009sr,Srednicki:1993im,2003JPhA...36L.205P}.  Using these techniques, the renormalized entanglement entropy was calculated numerically in~\cite{Liu:2012ee} for the free massive scalar. 

For the free massive scalar, one first expands the field into modes of integer angular momentum $n$.  The radial direction is discretized into $N$ units.  The discrete Hamiltonian for the $n^{\text{th}}$ angular momentum mode is given by $H_n = {1\over2} \sum_i \pi_i^2 + {1\over2} \sum_{ij} \phi_i K_n^{ij} \phi_j$, with $\pi_i$ the conjugate momentum to $\phi_i$ and $i = 1, \dots, N$.  The matrix $K_n$ has nonzero entries~\cite{Huerta:2011qi}   
\es{Kmat}{
K_n^{11} = {3 \over 2} + n^2 + m^2 \,, \qquad K_n^{ii} = 2 + {n^2 \over i^2} + m^2 \,, \qquad K_n^{i,i+1} = K_n^{i+1,i} = - {i+1/2 \over \sqrt{i(i+1)} } \,.
}
From here one constructs the two-point correlators $X \equiv \left< \phi_i \phi_j \right> ={1\over 2} (K^{-1/2})_{ij}$ and $P \equiv \left< \pi_i \pi_j \right> ={1\over 2} (K^{+1/2})_{ij}$.  To calculate the entanglement entropy across a circle of radius $R$ in lattice units, one must then reduce the matrices $X_{ij}$ and $P_{ij}$ to the $r \times r$ matrices $X^r_{ij}$ and $P^r_{ij}$, which are defined by taking  $1 \leq i,j \leq r$ with $r = R - {1\over2}$.  The entanglement entropy is then given by 
\es{NumS}{
S(R) = S_0 + 2 \sum_{n=1}^\infty S_n \,,
}
with~\cite{Huerta:2011qi} 
\es{Sn}{
S_n = \tr \left[ \left( \sqrt{X_n^r P_n^r} + \frac12 \right) \log \left( \sqrt{X_n^r P_n^r }+ \frac12 \right) - \left( \sqrt{X_n^r P_n^r} - \frac12 \right) \log \left( \sqrt{X_n^r P_n^r} - \frac12 \right) \right] \,.
}      

The calculation for the free massive Dirac fermion proceeds analogously to that for the scalar.  We write the two-component Dirac spinor as $\psi = (u,v)^{\text{T}}$.  We again discretize the radial direction into an $N$ unit lattice.  Expanding the field $\psi$ into modes of half-integer angular momentum $n$, we write the discrete Hamiltonian as 
\es{}{
H = \sum_{ij} \left( M_{ij}^{11} u_i^{\star} u_j + M_{ij}^{12} u_i^{\star} v_j + M_{ij}^{21} v_i^{\star} u_j + M_{ij}^{22} v_i^{\star} v_j \right) \,, 
}
where the 4 matrices $M^{11}$, $M^{12}$, $M^{21}$, and $M^{22}$ can conveniently be combined into the $2 N \times 2 N$ matrix $\tilde M_n^{2 k+\alpha - 2, 2 \ell +\beta - 2} = M^{\alpha \beta}_{k l}$, with $\alpha, \beta = 1,2$ and $k, l = 1, \dots, N$.  The nonzero entries of $\tilde M_n$ are~\cite{Huerta:2011qi}  
\es{Mmatrix}{
&\tilde M_n^{kk} = (-1)^{k+1} m \,, \qquad \tilde M_n^{12} = i \left( n + \frac12 \right) \,, \qquad \tilde M_n^{21} = -i \left( n + \frac12 \right) \,, \\
&\tilde M_n^{2k-1,2k} = i {n \over k} \,, \qquad \tilde M_n^{2k,2k-} =- i {n \over k} \,, \qquad  \tilde M_n^{2k-1,2k+2} =-{ i \over 2} \,, \\
&\tilde M_n^{2k,2k-3} ={ i \over 2} \,, \qquad \tilde M_n^{2k-1,2k-2} ={ i \over 2} \,, \qquad \tilde M_n^{2k,2k+1} =-{ i \over 2} \,.
} 

The two-point correlator $< \psi_i \psi_j^{\dagger} >$ for the $n^{\text{th}}$ angular momentum mode is then given by the matrix $C_n = \Theta(-\tilde M_n )$, which in practice can be calculated by first diagonalizing $-\tilde M_n$, taking the step-function of the eigenvalues, and then rotating back.  The contribution to the entanglement entropy of the $n^{\text{th}}$ mode is $S_n(R) = - \tr \left[ (1 - C_n^r) \log(1 - C_n^r) + C_n^r \log C_n^r \right]$, where $C_n^r$ denotes the restriction of $C_n$ to the first $(2r +1)$-dimensional block.  The entangling circle radius $R$ is related to $r$ by $R = r + {1 \over 2}$, and the total entanglement entropy is found by summing over the contributions of all the angular momentum modes: $S(R) ={1 \over 2} \sum_n S_n(R)$, with the factor of $1/2$ coming from the well known effect of fermion doubling on the radial lattice.   
 We refer the interested reader to~\cite{Huerta:2011qi} for further details of the numerical computation.

We discretize the radial direction into a lattice consisting of $N = 200$ points.  To minimize lattice effects, we only calculate the entanglement entropy for $30 < r < 50$ in lattice units.  For both the scalars and fermions, we calculate the entropy for $ m = .005\cdot i$ in inverse lattice units, with $i = 0, \dots, 16$, which approximately allows us to cover the range $.15 \leq (m R) \leq 4$.  Finite lattice effects are most important for the modes with small angular momentum~\cite{Lohmayer:2009sq}.  Following~\cite{Liu:2012ee}, for the lowest $10$ angular momentum modes we perform the numerical calculation on lattices consisting of $200+10\cdot j$ points, with $j = 0, \dots,10$, and then extrapolate to infinite lattice size.  As in~\cite{Huerta:2011qi}, we sum explicitly over the first $3000$ angular momentum modes and take into account higher angular momentum modes by fitting 
\es{sn}{
S_n = a_2 {1 \over n^2} + b_2 {\log n \over n^2} + a_4 {1 \over n^4} +b_4 {\log n \over n^4} +a_6 {1 \over n^6} +b_6 {\log n \over n^6} 
}
and then summing over the rest of angular momentum modes using the approximate fit function.            

\begin{figure}[htb]
  \leavevmode
\begin{center}$
\begin{array}{cc}
\scalebox{.9}{\includegraphics{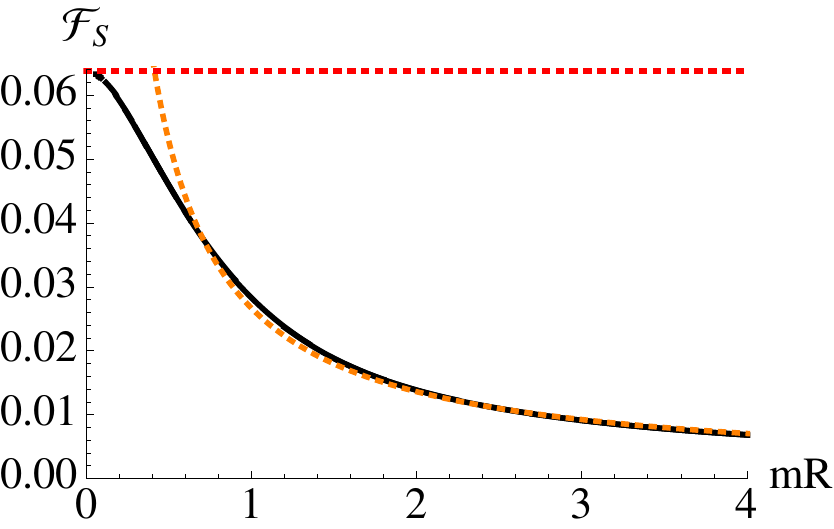}} & \scalebox{.9}{\includegraphics{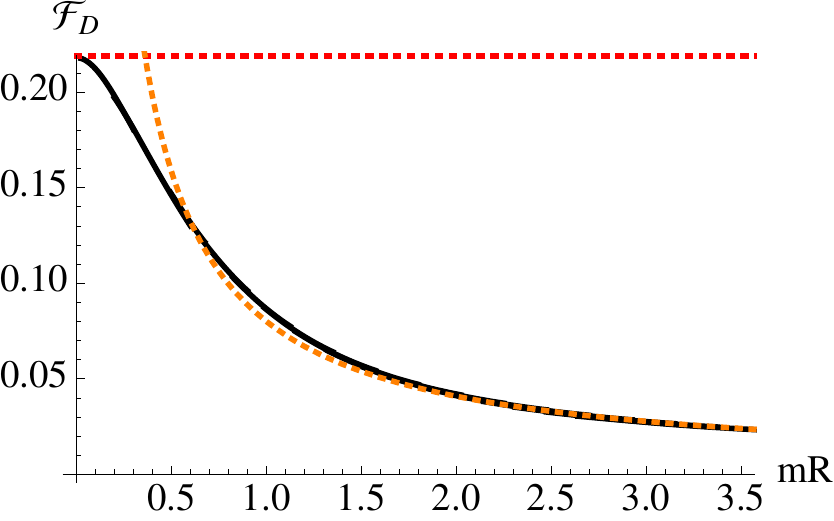}}
\end{array}$
\end{center}
\caption{The renormalized entanglement entropy ${\cal F}$ for the massive real free scalar ({\bf left}) and massive free Dirac fermion ({\bf right}).  The black curves are the results of the numerical lattice computations.  The orange curves are the analytic $1/(mR)$ IR approximations coming from the coefficient $\tilde c^{\Sigma}_{-1}$ in~\eqref{a2Rulekappa}.  The dotted red lines are the $m = 0$ values $F_S$ and $F_D$ for the scalar and fermion, respectively.}  
\label{numResults}
\end{figure}

In figure~\ref{numResults} 
 we plot the numerical results for the renormalized entanglement entropy for the massive scalar and Dirac fermion along with the analytic $1/(mR)$ approximation coming from the coefficient $\tilde c^{\Sigma}_{-1}$ in~\eqref{a2Rulekappa}.  The numerical results are in agreement with the analytic approximation.  The coefficient  $\tilde c^{\Sigma}_{-1}$ was checked in a similar way in~\cite{Huerta:2011qi}.  At $m = 0$, the renormalized entanglement entropy for the scalar and fermion should approach the values $F_S = (2 \log 2 - 3 \zeta(3) / \pi^2 )/2^4 \approx .0638$ and $F_D = (2 \log 2 + 3 \zeta(3) / \pi^2 )/2^3 \approx .2190$ in~\cite{Klebanov:2011gs}, respectively.  We are able to confirm this to high numerical accuracy.  This check was performed in~\cite{Liu:2012ee} for the massless scalar.     
 
\begin{figure}[htb]
  \leavevmode
\begin{center}
\scalebox{1.1}{\includegraphics{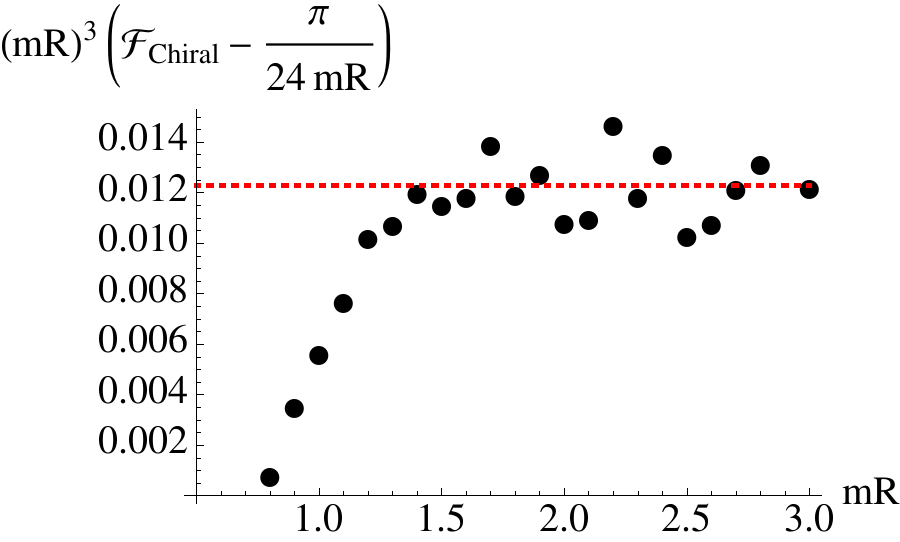}} 
\end{center}
\caption{The function $(mR)^3 \left( {\cal F}_{\text{chiral}} - {\pi \over 24} {1 \over mR} \right) $ plotted at large values of $(mR)$, where ${\cal F}_{\text{Chiral}}$ is the renormalized entanglement entropy of the free massive chiral multiplet, which consists of two real scalar fields and one Dirac fermion.  The black points  are the results of the numerical lattice computation, which is subject to numerical error at this level of precision.  The dotted red line is the predicted asymptotic value ${\pi \over 256}$, which we compute using~\eqref{FIRchiral}.}
\label{Fasymp}
\end{figure}

 We may test the $1/(mR)^3$ term in~\eqref{FIRchiral} by plotting the function 
 \es{Fasympf}{
 (mR)^3 \left( {\cal F}_{\text{chiral}} - {\pi \over 24} {1 \over mR} \right) = {\pi \over 256} + O\left(1 / (mR)^2 \right) 
 } 
at large values of $(mR)$, where ${\cal F}_{\text{chiral}}$ is the renormalized entanglement entropy of the massive free chiral multiplet.  In figure~\ref{Fasymp} 
we plot this function along with the hypothesized asymptotic value ${\pi \over 256}$.  While unfortunately numerical error does become significant at this level of precision, the numerical results do seem to be in agreement with the asymptotic value ${\pi \over 256}$ predicted in~\eqref{FIRchiral}.

  \section{Discussion} 
  
  There are a number of interesting ways in which our work could be generalized.  One could try and generalize~\eqref{Slog7Fin2} to allow for entangling surfaces for which $k^{0}_{ab} \neq 0$.  Another obvious way~\eqref{Slog7Fin2} could be generalized is to allow for theories which don't obey the constraint $B_3 = \frac13 \left(B_2 - {1 \over 2} B_1\right)$.  One way of doing this might be to use a more general gravitational theory.  However, it should also be possible to derive~\eqref{Slog7Fin2} without resorting to holography.
  
  Another interesting avenue would be to generalize the $(5+1)$-dimensional calculation of the entanglement entropy across $\Sigma = T^3 \times S^1$ to $(4+2 n)$-dimensions with $\Sigma = T^{1+ 2n} \times S^1$ and integer $n \geq 0$.  The resulting expression evaluated for a multiplet consisting of a free massless Dirac fermion and $2^{n+2}$ free real massless scalars would, in conjunction with~\eqref{acIdent}, give the $1/(mR)^{2n+1}$ term in the $(2+1)$-dimensional renormalized entanglement entropy of the massive free chiral multiplet.  Since the massive free chiral multiplet is perhaps the simplest massive theory in $(2+1)$-dimensions, it would be interesting to see if the expansion of its renormalized entanglement entropy re-sums in a simple way.  Carrying out this procedure would require calculating the conformal anomaly coefficients (or some combination thereof) of the multiplet consisting of a Dirac fermion and $2^{n+2}$ real scalars in  $(4+2 n)$-dimensions.
  
  It would also be interesting to connect these results with recent attempts~\cite{Elvang:2012st, Elvang:2012yc, Bhattacharyya:2012tc,Hoyos:2012xc} at proving the $6$-dimensional $a$-theorem~\cite{Cardy:1988cwa}.  These attempts all use the methods applied by Komargodski and Schwimmer in~\cite{Komargodski:2011vj,Komargodski:2011xv} to successfully prove the $4$-dimensional $a$-theorem.  It might be the case, however, that an entanglement entropy based proof the $6$-dimensional $a$-theorem, similar to the proofs given in $2$ and $3$-dimensions by Casini and Huerta~\cite{Casini:2004bw,Casini:2012ei}, is also possible and perhaps more tractable.        
  
  \section*{Acknowledgments}
We thank H.~Casini, M.~Huerta, L.Y.~Hung, R.~Myers, and A.~Singh for helpful discussions and correspondences, and especially S.~Pufu, T.~Nishioka, and I.R.~Klebanov for 
 involvement in early stages of this project, careful readings of this manuscript and numerous useful comments.  We also thank C.~Gerlein for technical assistance with the numerical computations. 
BRS was supported by the NSF Graduate Research Fellowship Program. BRS thanks the Institute for Advanced Study and the Perimeter Institute for hospitality.

\bibliographystyle{ssg}
\bibliography{lovelock2}

\end{document}